\documentclass[a4paper]{jpconf}

\usepackage[english]{babel}
\usepackage{graphicx}
\usepackage{wasysym,amssymb}
\usepackage[outercaption]{sidecap} 

\begin{document}

\title{Spinning black hole in the puncture method: Numerical experiments}
\author{Tim Dietrich, Bernd Br\"ugmann} 
\address{Friedrich-Schiller-Universit\"at Jena, 07743 Jena, Germany}
\ead{tim.dietrich@uni-jena.de, bernd.bruegmann@uni-jena.de}

\begin{abstract}
The strong-field region inside a black hole needs special attention during numerical simulation. 
One approach for handling the problem is the moving puncture method, which has become an important 
tool in numerical relativity since it allows long term simulations of binary black holes. 
An essential component of this method is the choice of the '1+log'-slicing condition. 
We present an investigation of this slicing condition in rotating black hole spacetimes. 
We discuss how the results of the stationary Schwarzschild '1+log'-trumpet change
when spin is added. This modification enables a simple and cheap algorithm for 
determining the spin of a non-moving black hole for this particular slicing condition. 
Applicability of the algorithm is verified in simulations of single black hole, binary 
neutron star and mixed binary simulations.
\end{abstract}

\section{Introduction}

The moving puncture method is a successful approach for the simulation of 
black holes, see \cite{Alcbook,BauShabook} for pedagogic explanations and references. 
Numerically the method can be applied to multiple, spinning black holes in quasi-circular 
or eccentric orbits, but analytical results are only known for the Schwarzschild spacetime. 
Without spin and linear momentum a single black hole settles 
down to a stationary state similar to a 'trumpet', containing one asymptotically flat end
and one with finite Schwarzschild radius, e.g.~\cite{HanHusPol06,Bro07,HanHusOhm08,Bru09}. 
In this article we begin an investigation of punctures in axisymmetry by performing 
numerical experiments of rotating black holes. 
We will show that the spinning puncture settles down to a stationary state similar to 
the spherical symmetric case and generalize results obtained for the Schwarzschild spacetime.
A short description about possible modification of the ansatzes of \cite{HanHusPol06} will be given, 
which offers the possibility to read off the spin value directly without additional computational costs.

\section{Spinning Puncture} \label{Spinning Puncture}

\subsection{Numerical setup}

We use the BAM-code \cite{BruGonHan06} for fully general 
relativistic evolutions with the moving puncture method \cite{CamLouMar06,BakCenCho06}. 
Unless otherwise stated the total mass of the system $M$ is set to unity and 
the spin axis is equivalent to the z-axis.  
We use the BSSN \cite{ShiNak95,BauSha99} evolution scheme for the numerical evolution. 
As a particular gauge choice, we apply the commonly used 1+log-slicing condition \cite{BonMasSei94}
together with the $\tilde{\Gamma}$-driver shift \cite{AlcBruDie04}.
This special choice, sometimes called moving puncture coordinates, is well established in
numerical relativity and widely used.

\subsection{Stationarity} \label{stationary}

\begin{figure}[t]
\begin{center}
\begin{minipage}[t]{0.45\textwidth}
\begin{center}
\includegraphics[width=1.0\textwidth]{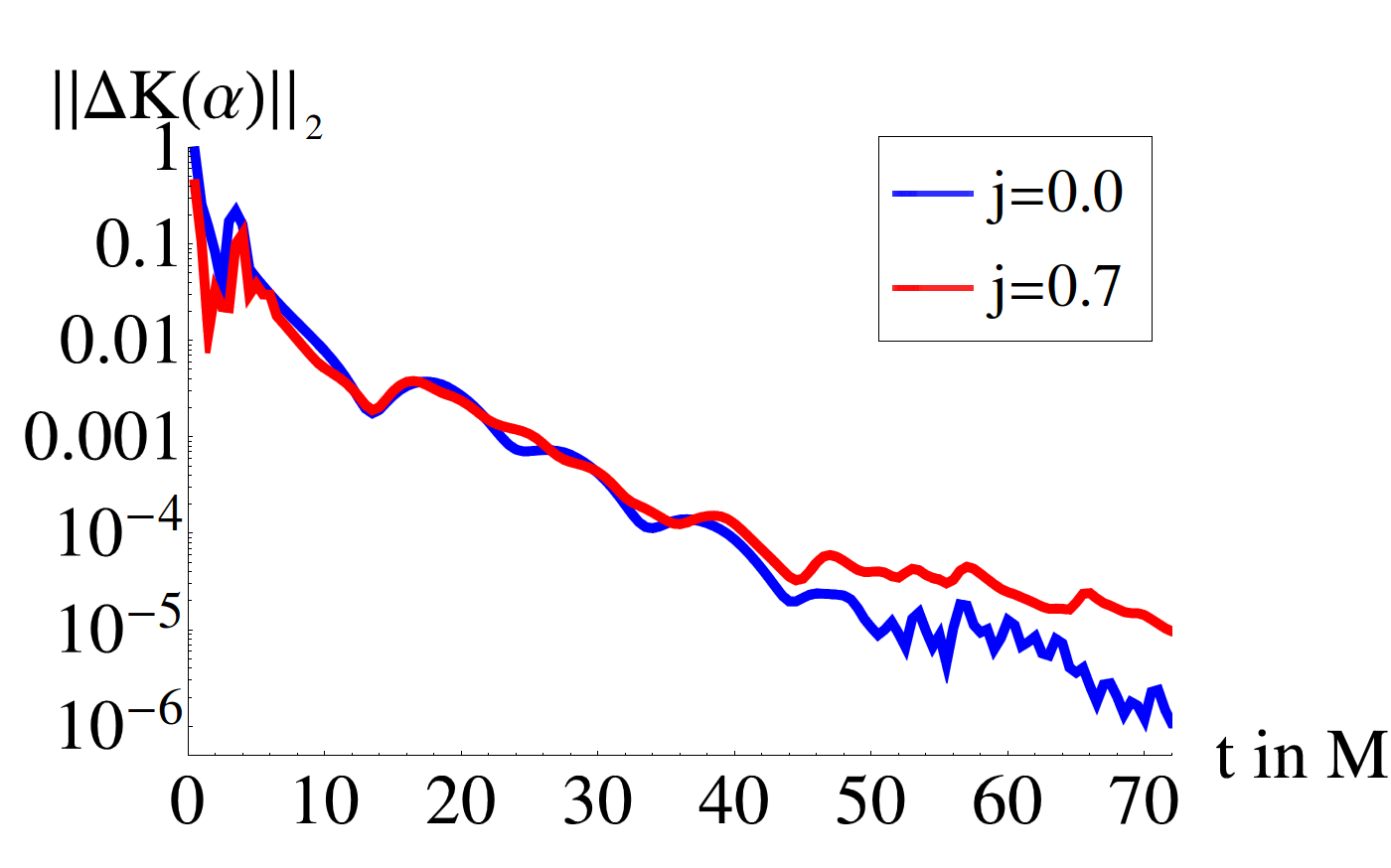} 
\caption{$||\Delta K(\alpha)||_2$ for $j=0.0$ and $j=0.7$ with $t_1-t_2=0.5M$.\label{picstat1}}
\end{center}
\end{minipage}
\hspace{\fill}
\begin{minipage}[t]{0.45\textwidth}
\begin{center}
\includegraphics[width=1.0\textwidth]{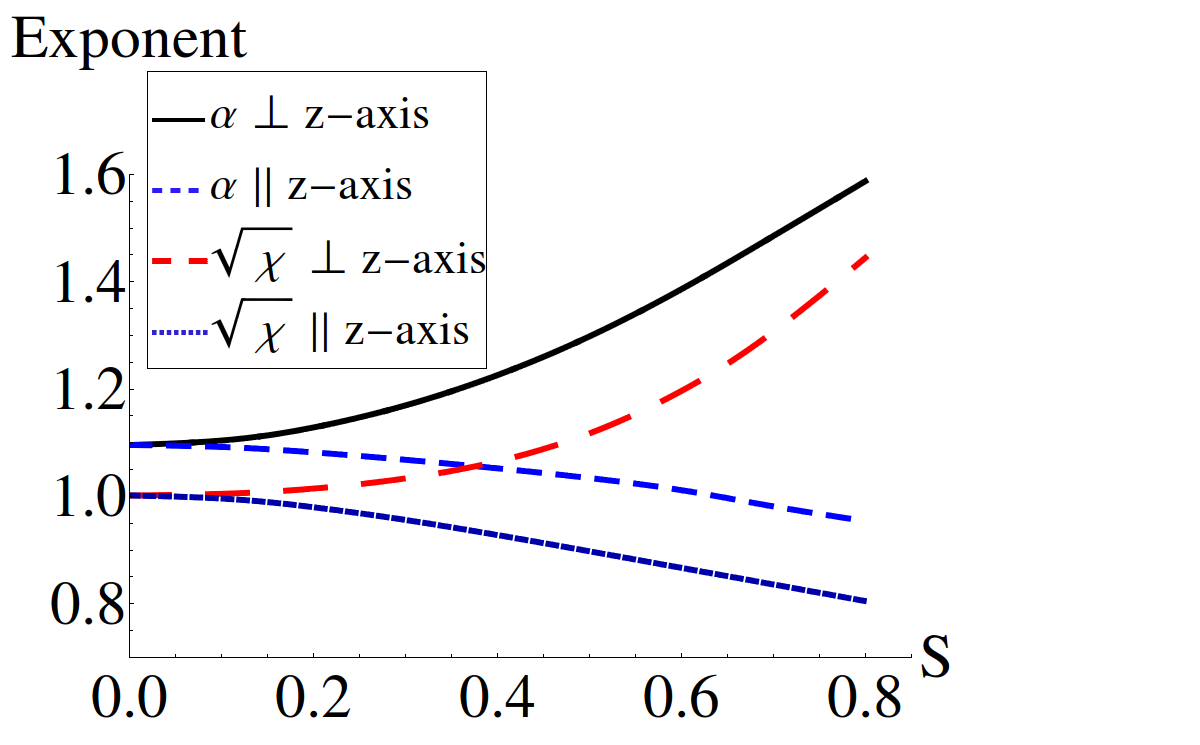}  
\caption{Exponents for the lapse and the conformal factor along different axes.\label{picexp}}
\end{center}
\end{minipage}
\end{center}
\end{figure}

As a starting point, we want to demonstrate that for spinning punctures a 
stationary state is reached similar to the one in spherical symmetry. For this purpose, 
we consider the trace of the extrinsic curvature $K$ as a function of the lapse $\alpha$, 
because $K(\alpha)$ 
is independent of the spatial coordinates as long as we use the same slicing condition for all simulations. 
When a stationary state is reached, 
the difference of $K(\alpha)$ between two hypersurfaces, i.e.\
$
||\Delta K(\alpha)||_2= 
\sqrt{\int | K(\alpha)_{t_1}-K(\alpha)_{t_2} |^2},
$
becomes zero for $t \rightarrow \infty$.
In Fig.~\ref{picstat1} we compare directly the spinning and the non-spinning configuration. 
We know that the non-spinning puncture reaches a stationary
state and the same seems to be true for the spinning black hole. 
Additionally, one can show that the time derivatives of the evolution variables 
become zero. We checked this for the lapse $\alpha$ by computing the right hand side of the 
1+log-slicing condition, e.g.~for a black hole with $j=0.4$ the derivative after $100M$
is smaller than $5 \cdot 10^{-5}$. 

\subsection{Influences on the BSSN-evolution variables}\label{BSSN}

The stationary state in spherical symmetry can be computed semi-analytically. 
The lapse behaves near the puncture like 
$\alpha \sim r^{1.0916}$, see \cite{Bru09}. We find an exponent of $\gamma = 1.091$ when we use 
our numerical data and apply the fitting function
$\alpha=\alpha_0+\alpha_1 r^\gamma  (1+\alpha_2 r +\alpha_3 r^2)$. 
We can do the same for the 
conformal factor $\psi$ and get for $\sqrt{\chi}=\psi^{-2}$ with 
$\sqrt{\chi}=c_0+c_1 r^\gamma  (1+c_2 r +c_3 r^2)$ an exponent of 
$\gamma=1.002$, while $\gamma=1.000$ is the analytical result. 
One should mention that we are using the 
radial coordinate of our numerical grid $r$, which does not necessarily agree 
with the isotropic radius for which the exponents were derived in \cite{Bru09}, 
but because of the particular gauge choices both descriptions become approximately the same
near the puncture. Thus, we will use the coordinate radius 
to get a better insight in the numerical simulation, not 
necessarily in the geometry of the spacetime.
According to Fig.~\ref{picexp} the exponents of $\alpha$ and $\sqrt{\chi}$ 
change differently, depending on the axis between the spin and the axis 
used for the fit. 

Remarkable is that even when the exponent 
is decreasing the puncture method works and can handle the non-differentiable 
functions. 
The important point is that although the regularity of the functions is reduced, 
compared with spherical symmetry, the moving puncture method does not fail.
This effect could explain why stationary spinning punctures are less accurate than 
non-spinning punctures in numerical simulations \cite{Lou08}.
Additionally, the angular dependence of the exponent $\gamma$ complicates possible 
attempts to factor out the singularity analytically. 

Independent of the non-integer powers of $r$ for $\alpha$ and $\sqrt{\chi}$, 
we can use a Taylor expansion to model the behavior near the black hole, see 
 \cite{HanHusPol06} for the description in spherical symmetry. 
Considering axisymmetry we have to introduce an additional angular dependence, 
where we define $\varphi$ as the angle between the axis of 
rotation $S^i$ and $n^i$ as the outward-pointing unit radial vector. 
This dependence is described reasonably 
well with spherical harmonics, e.g.~for the trace of the extrinsic curvature 
$K=\sum_{i=0}^{\infty} K_i r^i$, where we choose 
$K_i = \sum_{l=0}^{\infty} K_{il} Y_{2l,0} (\varphi,0)$ to encode axisymmetry in 
the components $K_i$. The same holds for the other BSSN scalars.
The shift vector $\beta^i$ in axial symmetry can be described by
$\beta^i = ( b_1 r + b_2 r^2) n^i + (c_1 r + c_2 r^2) \bar{\epsilon}^{ijk} n_j S_k,$
where we introduced the additional term $(c_1 r + c_2 r^2) \bar{\epsilon}^{ijk} n_j S_k$. 
A closed description of the extrinsic curvature $\tilde{A}_{ij}$ and the 
conformal metric $\tilde{g}_{ij}$ is more complicated and is explained 
up to some point in \cite{Die12}. 

\section{Spin-determination without integration}

\subsection{The new algorithm}\label{algorithm}

\begin{SCfigure}
\includegraphics[width=0.375\textwidth]{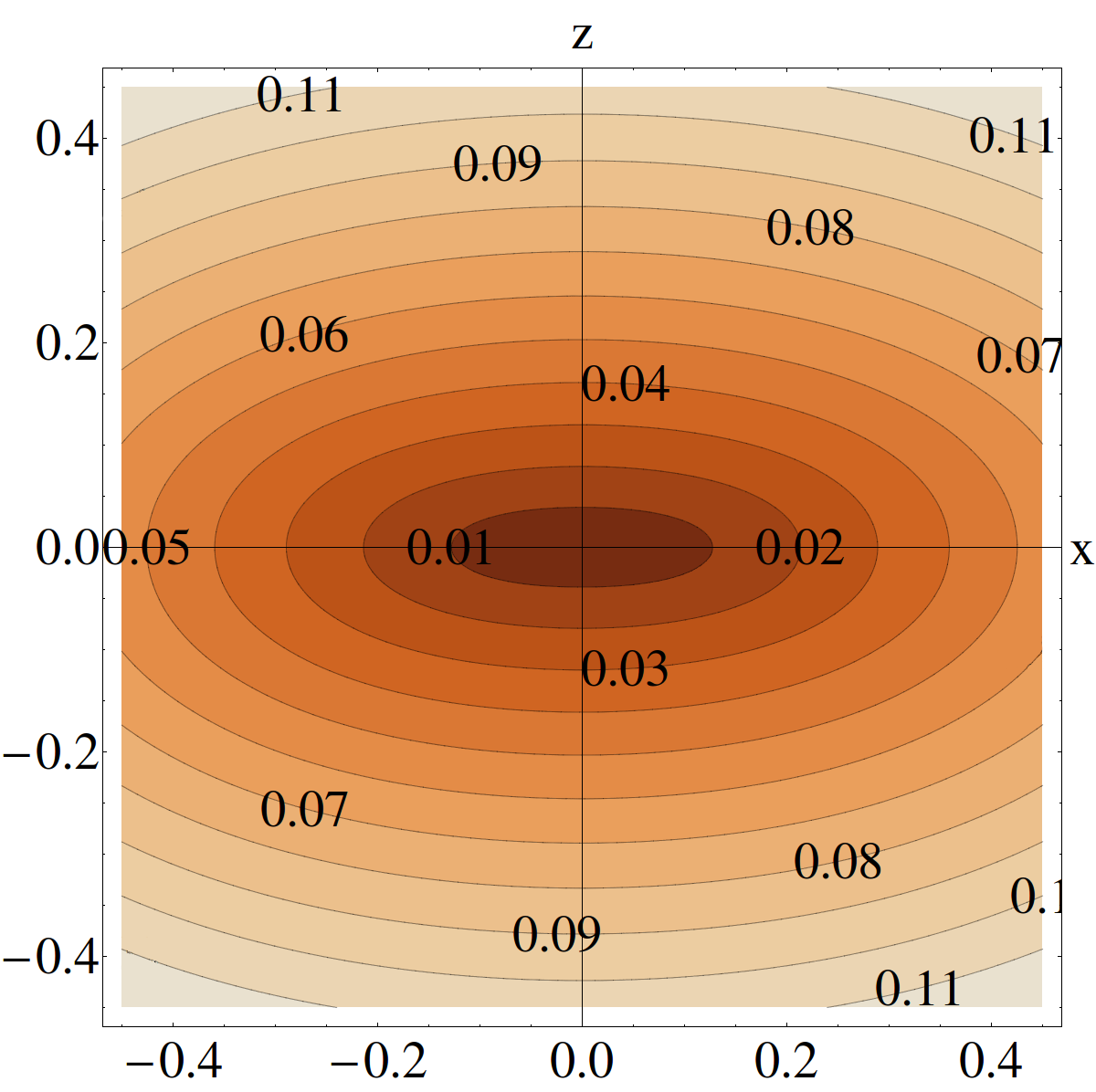} \qquad
\caption{ Level surface of $\alpha$ for a black hole with $j=0.8$. 
Although the contour lines might suggest that the puncture is stretched to a 
one dimensional line instead of a single point numerical experiments showed no indication that 
the puncture has an extension $> 10^{-3}M$. Thus, also a spinning puncture can be treated as 
a point.}
\label{plot:contour}
\end{SCfigure}

An important issue during the numerical evolution of black holes is obtaining a good estimate 
of the black hole spin. 
Using the properties of the stationary state 
described in sec.~\ref{stationary}, it is simple to find the spin of a 
non-moving black hole evolved with the puncture gauge. 
For increasing spin the absolute value of $K$ 
at the puncture, $K_0$, becomes smaller. The behavior is 
well approximated by a quadratic function. Inverting this formula leads to 
\begin{equation}
 S= \sqrt{1.41789 - 4.71218 K_0} \label{S(K)},
\end{equation}
which correlates an evolution variable of the BSSN scheme and properties 
of the underlying spacetime. Thus, we are able to read off the amplitude of the spin of 
a non-moving puncture simply by measuring $K_0$.
To get the entire information of the black hole spin we need 
the spin axis. This can be found with the lapse $\alpha$. Our experiments showed that
the coordinate distance between the origin and a contour 
line is smallest when we look along the rotation axis, see Fig.~\ref{plot:contour}.
The direction of rotation can be determined with the shift vector $\beta^i$, because of the 
additional term $(c_1 r + c_2 r^2) \bar{\epsilon}^{ijk} n_j S_k$. 
Empirical studies showed $c_1>0$, so we can find the sign of $S_k$ by regarding 
$\beta^i$ along one axis. 

\subsection{Numerical tests and measurement uncertainties} \label{Numerical tests}

The first test for the validity of our new algorithm are single black hole simulations.
Table~\ref{testcases} shows some test cases 
in which the spin was calculated using BAM's curvature flow apparent 
horizon finder and the new method. 
For further tests we include results for the final remnant of a mixed binary (BHNS) and 
a neutron star binary (NSNS) merger. Comparing with the apparent 
horizon finder, the results agree within the estimated uncertainty.

To define the measurement uncertainty we investigate every part of our algorithm individually. 
The first step is to determine the absolute spin value with the help of 
the interpolation function~(\ref{S(K)}). The error comes from the 
estimate of $K_0$ and can be evaluated by using different resolutions. 
Using the corresponding uncertainty, we can compute the error 
according to equation~(\ref{S(K)}), including also the standard error of the regression. 
The second step of our algorithm defines the angular uncertainty. 
Considering a box with $N^3$ grid points the angular error is 
approximately $2/N$, in our case $\lesssim 0.02 rad$. 
On the other hand using an interpolation to find the 
level surface one achieves a much lower uncertainty of around $\sim 10^{-3} rad$.
Because the last step of our algorithm uses only the qualitative behavior of $\beta^i$, 
it comes without an additional error. 
We can sum up that our method determines the spin with an uncertainty of 
$\Delta S \lesssim 0.02$ and an angular resolution of at least $0.02rad$ 
for $j \gtrsim  0.2$.
For smaller spins the change of $S(K)$ in equation (\ref{S(K)}) is small, which 
leads to higher uncertainties.  
Furthermore, equation (\ref{S(K)}) was defined as an adequate function for $j \in [0,0.7]$, 
nevertheless using higher resolution for higher spins the method is applicable as well.

 \begin{table}[htpb]

\caption{Single black hole simulations (left) and
results for the final remnant for a binary merger (right). 
The first line refers to the new algorithm, 
the second line to the apparent horizon finder.}

 \begin{tabular}{l l}

 \begin{small}
 \begin{tabular}{ c | c c c c c }
BH-mass  & $S_x$ & $S_y$ &  $S_z$ & j\\
\hline
\hline
  1 & 0.10 & 0.66 & 0.21 & 0.70$\pm$0.01 \\
    & 0.11 & 0.66 & 0.20 & 0.70 \\  
\hline
  1 & 0.09 & -0.50 & 0.21 & 0.55$\pm$0.01 \\
    & 0.10 & -0.50 & 0.20 & 0.55\\  
\hline
 1.92     & 0.04 & -0.32 & 0.15 & 0.10$\pm$0.07\\
          & 0.10 & -0.50 & 0.20 & 0.15\\    
\end{tabular}

\hspace*{\fill}

\begin{tabular}{ c | c c c }
  system  & $M_{ADM}^{init}$ & $K_0$ &$j$ \\
\hline
\hline
  BHNS & $8.328 M_{\astrosun}$ & $\sim 0.0315 M_{\astrosun}$ & $0.42 \pm 0.05$  \\
       & & & $0.43$\\
\hline
  NSNS & $2.998 M_{\astrosun}$ & $\sim 0.059 M_{\astrosun}$ &  $0.76 \pm 0.04$ \\
       & & & $0.76$ \\
\hline
  NSNS & $3.439 M_{\astrosun}$ & $\sim 0.057 M_{\astrosun}$ &  $0.71 \pm 0.05$ \\
       & & & $0.70$\\
\end{tabular}
\end{small}
 \end{tabular}
\label{testcases}
\end{table}

\section{Conclusion}

We presented numerical results for rotating black holes and generalized 
the ansatzes of \cite{HanHusPol06}. For spinning punctures
a simple $r^{\gamma(\varphi)}$-model for the evolution variables is possible, where in contrast 
to spherical symmetry the exponent depends on the angle $\varphi$. 
This complicates attempts to factor out the singularity analytically. In the future
we plan to use these results to find an analytical description for a rotating black hole using the 
moving puncture gauge. 

Additionally, a fast algorithm was introduced to determine the spin of a non-moving puncture. 
Adding linear momentum to a spinning black hole to obtain a binary orbit breaks the 
axisymmetry, but a generalization of the numerical method appears possible.

\section*{Acknowledgement}

It is a pleasure to thank D. Hilditch and G. Loukes-Gerakopoulos for helpful discussions and comments.
This work was supported in part by DFG grant SFB/Transregio~7
``Gravitational Wave Astronomy'',  the
Graduierten-Akademie Jena, and the Studienstiftung des deutschen
Volkes. Computations were performed at the Quadler cluster of the
Institute of Theoretical Physics of the University of Jena
and on SuperMUC of the Leibniz Rechenzentrum.

\end{document}